\documentclass{llncs}
\pdfoutput=1
\usepackage{epsfig}
\usepackage[latin1]{inputenc}
\graphicspath{{Figures/}}

\title{RapidMind: Portability across Architectures and its Limitations}
\author{Iris Christadler \and Volker Weinberg}
\institute{Leibniz-Rechenzentrum der Bayerischen Akademie der Wissenschaften,\\
  D-85748 Garching bei M\"unchen, Germany}

\begin{document}

\maketitle
\begin{abstract}
Recently, hybrid architectures using accelerators like  GP\-GPUs or the Cell processor have gained much interest in the HPC community. The ``RapidMind Multi-Core Development Platform'' is a programming environment that allows generating code which is able to seamlessly run on hardware accelerators like GPUs or the Cell processor and multi-core CPUs both from AMD and Intel. This paper describes the ports of three mathematical kernels to RapidMind which have been chosen as synthetic benchmarks and representatives of scientific codes. Performance of these kernels has been measured on various RapidMind backends (cuda, cell and x86) and compared to other hardware-specific implementations (using CUDA, Cell SDK and Intel MKL). The results give an insight into the degree of portability of RapidMind code and code performance across different architectures. 
\end{abstract}

\section{Introduction}
The vast computing horsepower which is offered by hardware accelerators and
their usually good power efficiency has aroused interest of the high
performance computing community in these devices. The first hybrid system
which entered the Top500 list~\cite{biba} was the TSUBAME cluster at Tokyo
Institute of Technology in Japan. Several hundred Clearspeed cards were used
to accelerate an Opteron based cluster; the system was ranked No. 9 in the
Top500 list in November 2006. Already in June 2006, a sustained Petaflop/s
application performance was firstly reached with the RIKEN MD-GRAPE 3 system
in Japan, a special purpose system dedicated for molecular dynamics
simulations. In 2008, the first system ever to reach a sustained High
Performance LINPACK (HPL) performance of more than one Petaflop/s was
``Roadrunner'', the No. 1 system on the lists in July 2008 and November
2008. Roadrunner is a hybrid system based on Opteron processors and accelerated with
PowerXCell8i processors, a variant of the Cell B.E. (Broadband Engine) with increased double-precision capability. 

However, applicability of hardware accelerators for general-purpose HPC systems is still a source of debate. In 2008, the landscape was quite diverse; many different hardware solutions existed (Cell, Nvidia and AMD/ATI GPUs, ClearSpeed accelerator boards, FPGA based systems) and every system had its own programming language and paradigm. At the same time, the x86 processors started to become multi-core processors and first HPC systems were based on hundred thousands of cores. Improving the scalability of HPC codes to be able to utilize the increased core counts was already difficult for the scientific communities; trying to add support for one of the new accelerators was a huge porting effort with a high risk: what if either the hardware or the software would not be supported on the long run? Solutions which offered support for different hardware architectures became appealing. 

While in the meantime several solutions (e.g. OpenCL~\cite{bib1}, PGI accelerator compiler~\cite{bib2}, CAPS hmpp~\cite{bib3}, StarSs~\cite{bib4}) exist which provide an abstraction of the underlying hardware characteristics from the programmer, the situation was different two years ago: RapidMind Inc. was one of the first companies providing support for general purpose computing on graphic processing units, nowadays known as GPGPUs. RapidMind started in 2004 based on the academic research related to the Sh project~\cite{bib12} at the University of Waterloo. Their work was started at a time when the first ``programmable'' GPUs were just released and the only way to program these devices was by using ``shading languages''. Already at that time people tried porting simulation codes to GPUs~\cite{bibFrank}. Since then, RapidMind has subsequently added the Cell processor backend (2007) and the x86 multi-core processor backend with the rise of multi-core processor CPUs for the consumer market (2008). In 2009, version 4.0 was released which introduced the cuda backend, necessary to support double-precision arithmetic on GPUs. 
Even today, RapidMind is still the only product that fully supports Cell, GPUs and multi-core CPUs. All other solutions are either limited by the hardware which they support or require an adaptation of the code.

At SC06 a paper was published which showed impressive performance gains by using RapidMind for porting three algorithms (SGEMM, FFT and Black-Scholes) to the GPU~\cite{bib7}. This is a follow-on work assessing the state-of-the-art three years later. However, the main reason for choosing RapidMind for a deeper investigation has been its programming paradigm which differs from serial programming languages and abstracts the massive parallelism of the underlying hardware more than any other language concept currently discussed for HPC.

\section{Overview}
\subsection{Software}

\begin{figure}[h]
\begin{center}
\begin{tabular}{cc}\fbox{\parbox[b]{4.5cm}{\tt\scriptsize
\medskip
\#include <rapidmind/platform.hpp>
using namespace RapidMind;\\
...      \\
// declaration\\
Array<1, Value4i> input;\\
Array<1, Value4f> output;\\

Program example = BEGIN \{\\
\hspace*{0.3cm}     // program definition\\
\} END;\\
// program call\\
output = example(input);
\medskip
  }}
&\epsfig{file=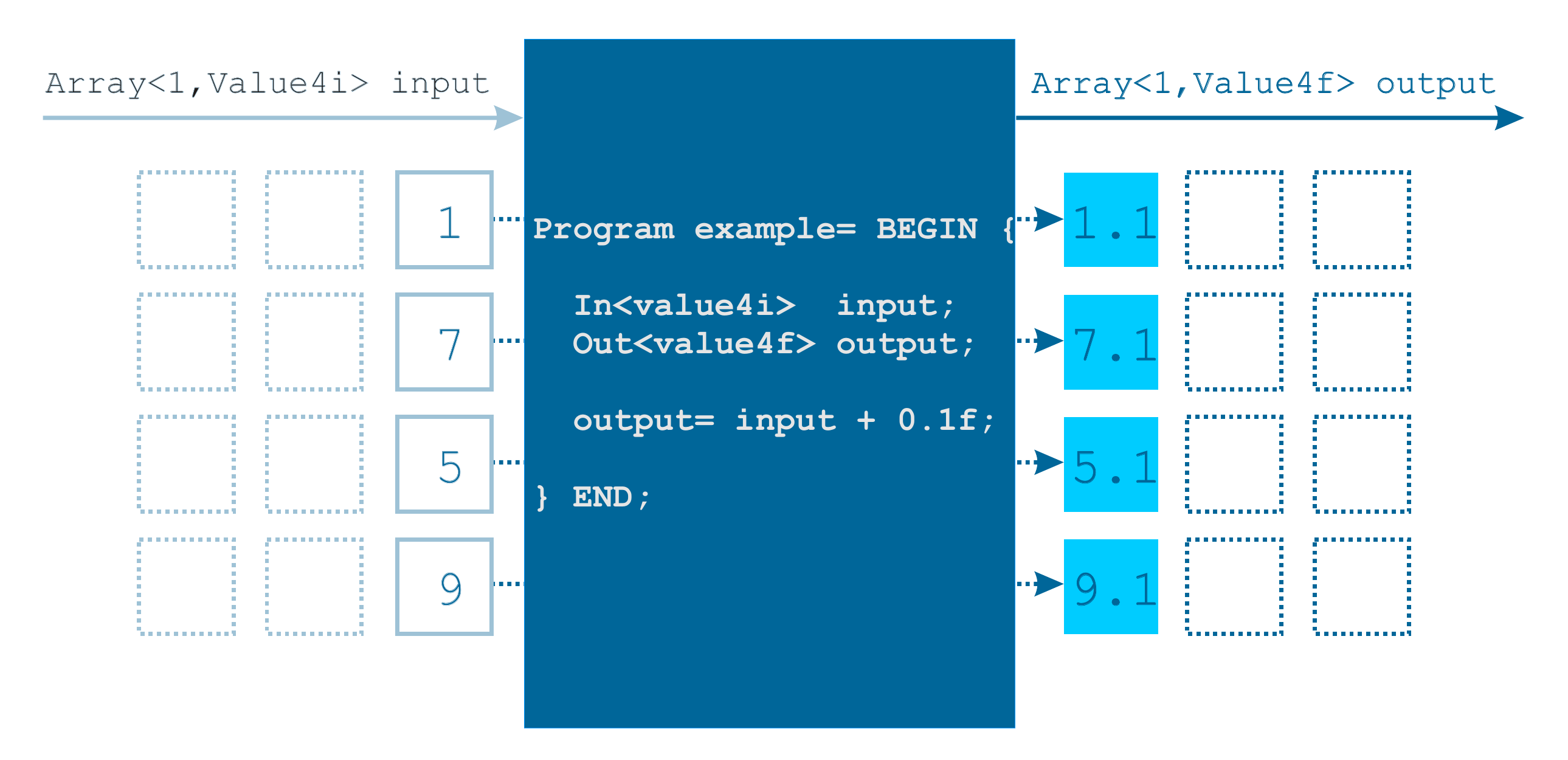,width=7.5cm}\\
(a) & (b)
\end{tabular}
\end{center}
\caption{RapidMind programming scheme}
\label{fig:RapidMind}
\end{figure}

The ``RapidMind Multi-Core Development Platform'' promises easy and portable access not only to multi-core chips from Intel and AMD but also to hardware accelerators like GPUs and Cell. The basic concept of the RapidMind language is called ``data-stream processing''; a powerful technology to express data parallelism. A simple example of a RapidMind program is given in Fig.~\ref{fig:RapidMind} (a).  Figure~\ref{fig:RapidMind} (b) represents a schematic view of the executed stream program.  A call to a RapidMind program can be inserted in any valid C++ program and needs to include the RapidMind library in the header of the file and during linkage. Unless specified explicitly, the RapidMind runtime environment will automatically search for available accelerator hardware and compile the program at runtime using the RapidMind backend for the detected hardware. 

RapidMind adds special types and functions to C++ which allow the programmer to define sequences of operations (RapidMind programs) on streams of data (special arrays). With these, data dependencies and data workflows can be easily expressed and will naturally contain all necessary information for an efficient (data-) parallelization. The compiler and the runtime environment then have maximum information to decide how to auto-parallelize the code.

The structure of RapidMind code forces the programmer to decide early in the development process which operations could be performed in parallel without any side-effects. This usually results in many small code snippets that can run in parallel which is optimal to fill the pipeline of a GPU or other massively parallel devices.

\subsection{Hardware}

\begin{table}[h]
\begin{center}
\begin{tabular}{l|rl|rl}\hline
\textbf {Hardware}&\multicolumn{2}{l|}{\textbf{SP peak perf.}}& \multicolumn{2}
{l}{\textbf{DP peak perf.}}\\\hline
1 C1060 GPU &    933 &GFlop/s & 78 &GFlop/s\\
1 Tesla S1070  &  4140 &GFlop/s & 345 &GFlop/s\\\hline
Nehalem-EP (2.53 GHz, 1 core)&20 &GFlop/s & 10 &GFlop/s\\
Nehalem-EP (2.53 GHz, 8 cores)& 162& GFlop/s    &81 &GFlop/s\\\hline

1 PowerXCell8i (8 SPUs) & 205 &GFlop/s  & 102 &GFlop/s\\
1 QS22-blade 2 PowerXCell8i (16 SPUs) & 410 &GFlop/s & 205 &GFlop/s\\\hline
\end{tabular}
\end{center}

\caption{Hardware overview}
\end{table}

Three different platforms are used for the performance measurements. An Nvidia Tesla based system is used to measure the cuda backend from RapidMind against implementations based on CUDA and the CUDA libraries cuBLAS and cuFFT. Tesla is Nvidia's first dedicated general purpose GPU with enhanced double-precision capability. 
A C1060 supports partly IEEE-754, consists of 240 thread processors with an overall performance of 78 GFlop/s in double-precision and 933 GFlop/s in single-precision. 
One Nvidia Tesla S1070 1U rack consists of four C1060 computing processors with a total single-precision performance of around 4 TFlop/s. 

An IBM QS22-blade based system is used to compare RapidMind's cell backend with code using Cell intrinsics which is taken from the SDK. Each QS22-blade hosts two PowerXCell8i, the processors used to accelerate Roadrunner~\cite{roadrunner}. Each PowerXCell8i is running at 3.2 GHz, is partly IEEE-754 conform and has a single-precision peak performance of 204.8 GFlop/s and a double-precision peak performance of 102.4 GFlop/s. A QS22-blade has therefore a total of slightly more than 400 GFlop/s single-precision performance. The main difference between the Cell processor and GPUs or current multi-core CPUs is its inhomogeneity; eight synergistic processor units (SPUs) are added to one PowerPC processor unit (PPU). The Cell processor has a very good performance per Watt ratio and the 6 most energy efficient supercomputers, as ranked by Green500~\cite{green500} in November 2009, are based on PowerXCell8i technology.

RapidMind's x86 backend is benchmarked against code using Intel's Math Kernel Library (MKL) on one of the latest Intel processors, a Xeon E5540 known as ``Nehalem-EP''. A Nehalem-EP core running at 2.53 GHz has a single-precision peak performance slightly above 20 GFlop/s and a double-precision peak performance of around 10 GFlop/s. One Nehalem-EP node consists of 2 sockets with four cores per socket. A Nehalem-EP node with 8 cores reaches 162 GFlop/s single and 81 GFlop/s double-precision performance. 

The performance figures of all three architectures are summarized in Table~1. Since the double-precision peak performance of one Nehalem-EP node (8 cores, 81 GFlop/s) is quite comparable with the double-precision performance of 1 Nvidia C1060 GPU (78 GFlop/s) and 1 PowerXCell8i (102 GFlop/s) we tried to compare these platforms directly where possible. 

\section{The RapidMind ports and their performance}

To judge the suitability of recent accelerator hardware for scientific
computing and high-performance computing, three mathematical kernels from the
Euroben benchmark suite~\cite{euroben} have been chosen: 
\begin{itemize}
\item	mod2am: a dense matrix-matrix multiplication,
\item	mod2as: a sparse matrix-vector multiplication,
\item	mod2f: a one-dimensional Fast Fourier Transformation (FFT).
\end{itemize}
The kernels have been selected to show both the advantages and the pitfalls of current accelerators. They are representatives of three (dense linear algebra, sparse linear algebra and spectral methods) of the ``seven dwarfs'', an ontology for scientific codes introduced by~\cite{berkeley}. According to Fig.~11 in~\cite{D61} these three dwarfs 
account for approximately one third of the workload of current European HPC Tier-1 centers. The selection of kernels was performed by the EU FP7-project PRACE, published in~\cite{D66} and should be extended to cover all important dwarfs in the future.

\subsection{Dense matrix-matrix multiplication (mod2am)}

The dense matrix-matrix multiplication ($C = A \times B$) is one of the most basic algorithms used in scientific computing. It is the basis of the High Performance LINPACK  code, which determines the Top500 rank of a system. The schoolbook version of the algorithm is composed of three nested for-loops.
Many sophisticated optimization strategies exist, and one of the fastest implementations is the MKL version. Making use of the MKL functions is straightforward and basically needs a call to \texttt{cblas\_dgemm} (double-precision arithmetic) or \texttt{cblas\_sgemm} (single-precision arithmetic). 

A first implementation in RapidMind is straightforward. In a first step, the RapidMind data types must be used to express the matrices $A$ (of size $m \times l$), $B$ ($l \times n$) and $C$ ($m \times n$). All matrices can be represented by two-dimensional arrays of floating point data:

\begin{verbatim} 
Array<2,Value1f> A(m,l);
Array<2,Value1f> B(l,n);
Array<2,Value1f> C(m,n);
\end{verbatim}

In a second step the RapidMind program \texttt{mxm} needs to be declared. Since there are no clear data streams which could be fed into the program a helper index array is used. This index array ensures that the defined program \texttt{mxm} can sum up the corresponding entries of the input matrices A and B. 
All matrices are automatically transferred to the GPU memory at execution time. The RapidMind control flow construct \texttt{RM\_FOR} is used to allow manipulation of the streamed data.

\begin{verbatim}
Program mxm = BEGIN {
    In<Value2i> ind;
    Out<Value1f> c = Value1f(0.);
    
    Value1i k;
    // Computation of C(i,j)
    RM_FOR (k=0, k < Value1i(l), k++) {
        c += A[Value2i(ind(0),k)]*B[Value2i(k,ind(1))];   
    } RM_ENDFOR;
} END;
\end{verbatim}


The call to the RapidMind program then looks as follows:

\begin{verbatim} 
C= mxm(grid(m,n));
\end{verbatim}

The call to the RapidMind function \texttt{grid(m,n)} generates a virtual helper array of size $m \times n$ which does not require additional storage. The whole helper array is automatically initialized with integers from $(0,0)$ to $(m,n)$ and is directly passed to the program and used for the index computation.

After this first naive approach a second, \emph{GPU-optimized version} of the matrix multiplication has been produced. This version is based on code available at the RapidMind developer portal~\cite{bib11}. The basic difference between both versions is the fact, that the GPU-optimized version operates on arrays of \texttt{Value4f}, to optimal use the GPU vector registers; $4 \times 4$ submatrices are multiplied and accumulated. 

\begin{figure}
\begin{center}
\includegraphics[width=0.80\textwidth]{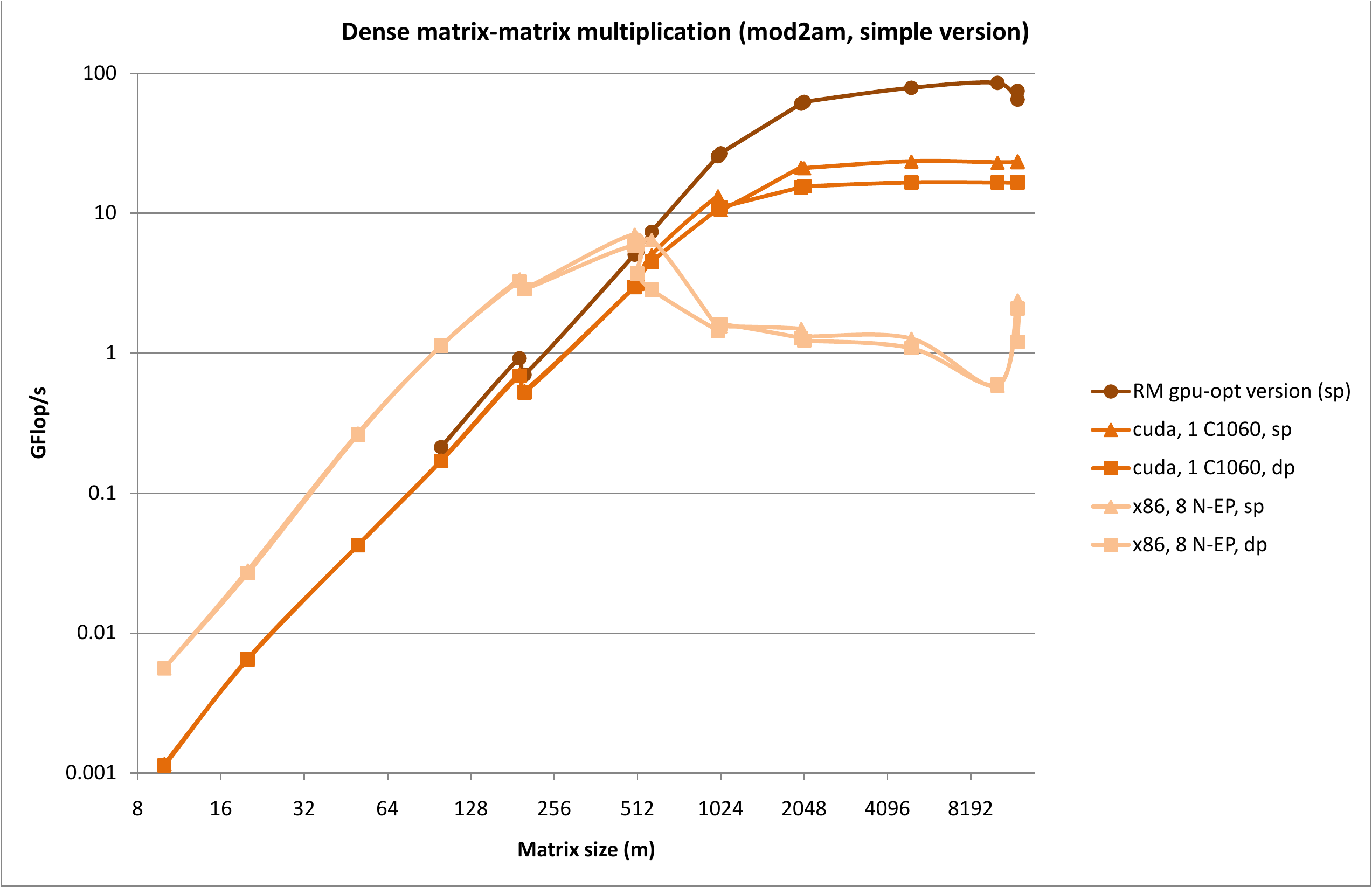}
\end{center}
\caption{Performance comparison of the simple mod2am version on various RapidMind backends and associated hardware. The simple version is also compared to the GPU-optimized version running on 1 C1060 GPU in single-precision.
}
\label{fig:mod2amSIMPLE}
\end{figure}

\begin{figure}
\begin{center}
\includegraphics[width=0.80\textwidth]{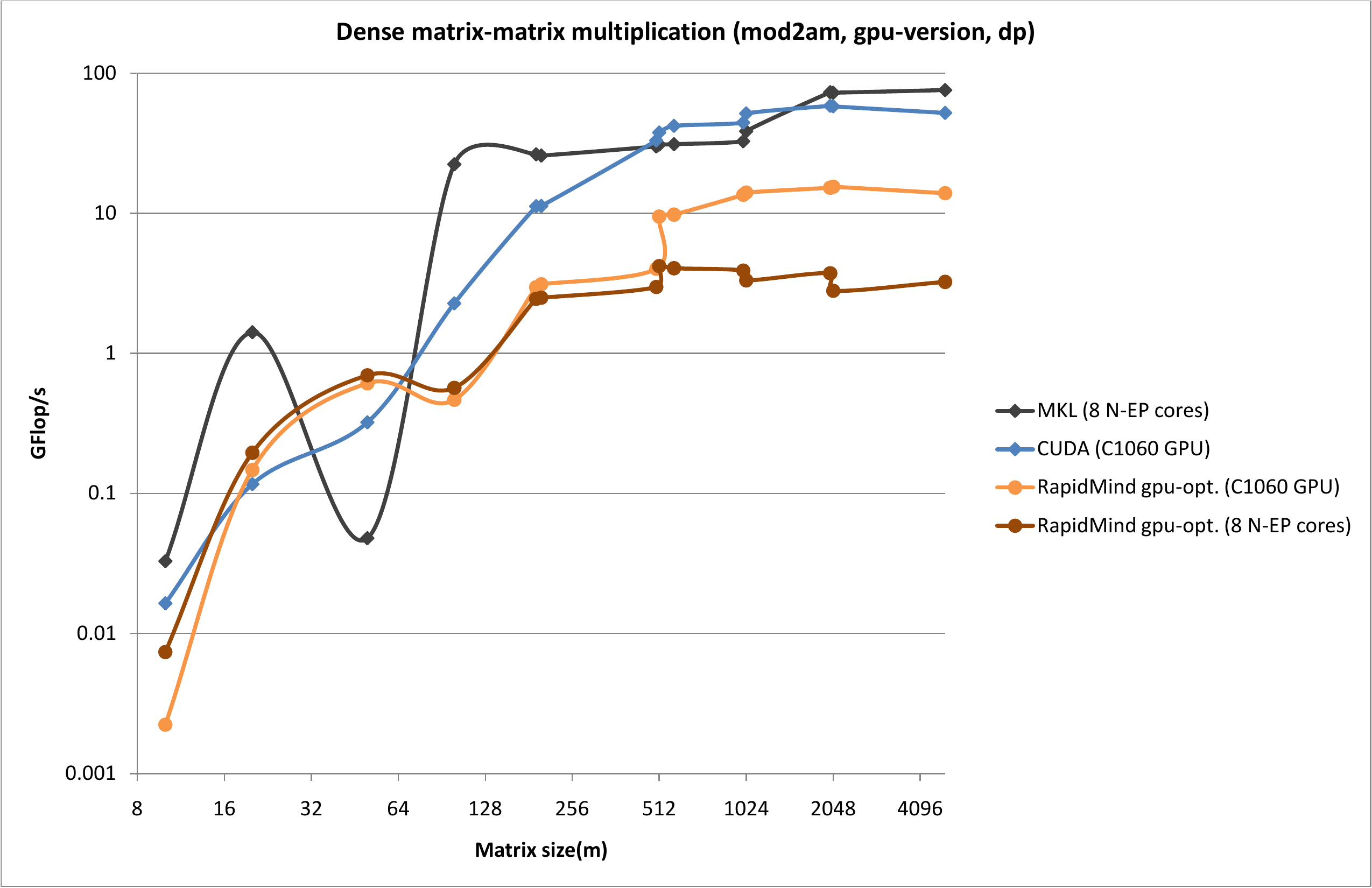}
\end{center}
\caption{Performance comparision of the GPU-optimized version on various backends. Performance measurements have been performed both on an Nvidia GPU and a Nehalem-EP socket with eight cores. The RapidMind version is compared to a CUDA version based on cuBLAS and an MKL implementation of the dense matrix-matrix multiplication. Performance measurements are based on double-precision arithmetic.}
\label{fig:mod2amGPU}
\end{figure}

\begin{figure}
\begin{center}
\includegraphics[width=0.80\textwidth]{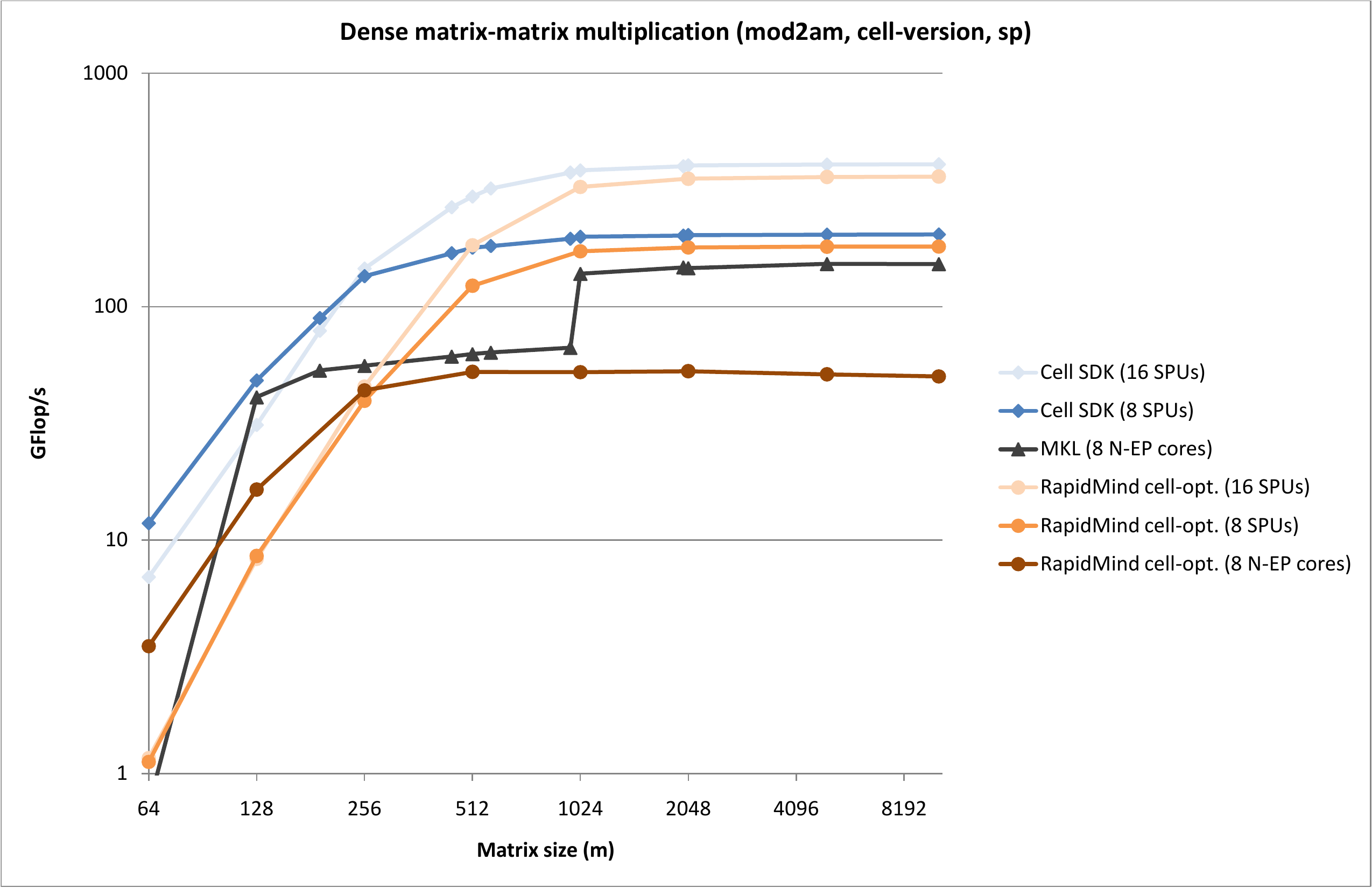}
\end{center}
\caption{Performance comparison of the Cell-optimized version on various RapidMind backends. Performance measurements have been performed on 1 PowerXCell8i (8 SPUs), 1 QS22-blade (16 SPUs) and an 8-core Nehalem-EP node. The RapidMind version is compared to a dense matrix-matrix multiplication example from the Cell SDK and the MKL implementation. All performance results are based on single-precision arithmetic.}
\label{fig:mod2amCELL}
\end{figure}

Figure~\ref{fig:mod2amSIMPLE} 
\footnote{Time is always measured for the execution of the whole kernel. This includes the time to transfer data between host and accelerator for GPU and Cell results. 
The y-axis uses log-scale to better cover the whole performance range for all matrix sizes.} 
shows the performance of the simple version using the cuda and x86 RapidMind backends and compares the single-precision cuda backend performance with the GPU-optimized version. 
It can be seen that the GPU-optimized version is indeed four times faster than the naive approach for single-precision arithmetic. This means that the use of \texttt{Value4f} instead of \texttt{Value1f} really improves performance. It is important to note, that neither the language nor the program definition of the simple approach should prevent the compiler from doing this optimization by itself. 

Measurements for double-precision reveal that the simple approach is actually faster than the GPU-optimized version. This is counterintuitive and only becomes understandable if one takes into account, that the cuda backend is the latest RapidMind backend and was introduced with version 4.0 in July 2009. The target of this version was to enable RapidMind to support Nvidia Tesla cards; a RapidMind version with improved performance of the cuda backend was scheduled for version 4.1.

Figure~\ref{fig:mod2amGPU} shows the performance of the GPU-optimized version on various backends and compares it with hardware-specific languages (CUDA and MKL).
It shows that the performance of the RapidMind implementation is more than an order of magnitude slower than the MKL implementation, while the difference between RapidMind and CUDA performance is only a factor of~3. Comparing Fig.~\ref{fig:mod2amSIMPLE} with Fig.~\ref{fig:mod2amGPU} reveals that the performance difference between the two RapidMind implementations varies extremely for certain matrix sizes, although the implementations vary only slightly. 

The performance of both the simple version and the GPU-optimized version are not able to deliver decent performance on the Cell platform. A third version \emph{optimized for the Cell processor} is based on another code available through the RapidMind developer portal. This time, computation is performed using a block partitioning of 64 by 64 blocks. All matrices are in a ``block swizzled'' format so that these blocks are contiguous in memory. The computations and memory transfers are overlapped using double buffering and are partly based on the matrix-matrix multiplication example from the IBM Cell SDK (\texttt{/opt/cell/sdk/src/demos/matrix\_mul/}). The Cell SDK version is also used for performance comparison. 

Figure~\ref{fig:mod2amCELL} gives an insight into the performance of the Cell-optimized version. 
Again the RapidMind figures have been compared with implementations in other languages. Since the Cell SDK version is based on single-precision arithmetic, it has been compared to single-precision results obtained with the RapidMind cell and x86 backends and an SGEMM implementation using MKL on 8 Nehalem-EP cores. 
This time, the RapidMind version is able to nearly meet the performance of the hardware-specific and highly optimized Cell SDK version; it reaches 88\% of the SDK version. However, this comes at the price of a hardware-specific RapidMind implementation and contradicts the idea of seamlessly portable code. 

In conclusion, the three different implementations illustrate the current limitations of code and performance portability. Hardly any problems were experienced when moving the code to other platforms, but in many cases the performance was not predictable. Tuning the code to better exploit certain characteristics of the employed hardware normally yields better performance but requires to stick with this hardware. The idea behind RapidMind is that the language and program definitions are generic enough to allow the compiler to do hardware-specific optimizations itself.

\subsection{Sparse matrix-vector multiplication (mod2as)}

\begin{figure}
\begin{center}
\includegraphics[width=0.80\textwidth]{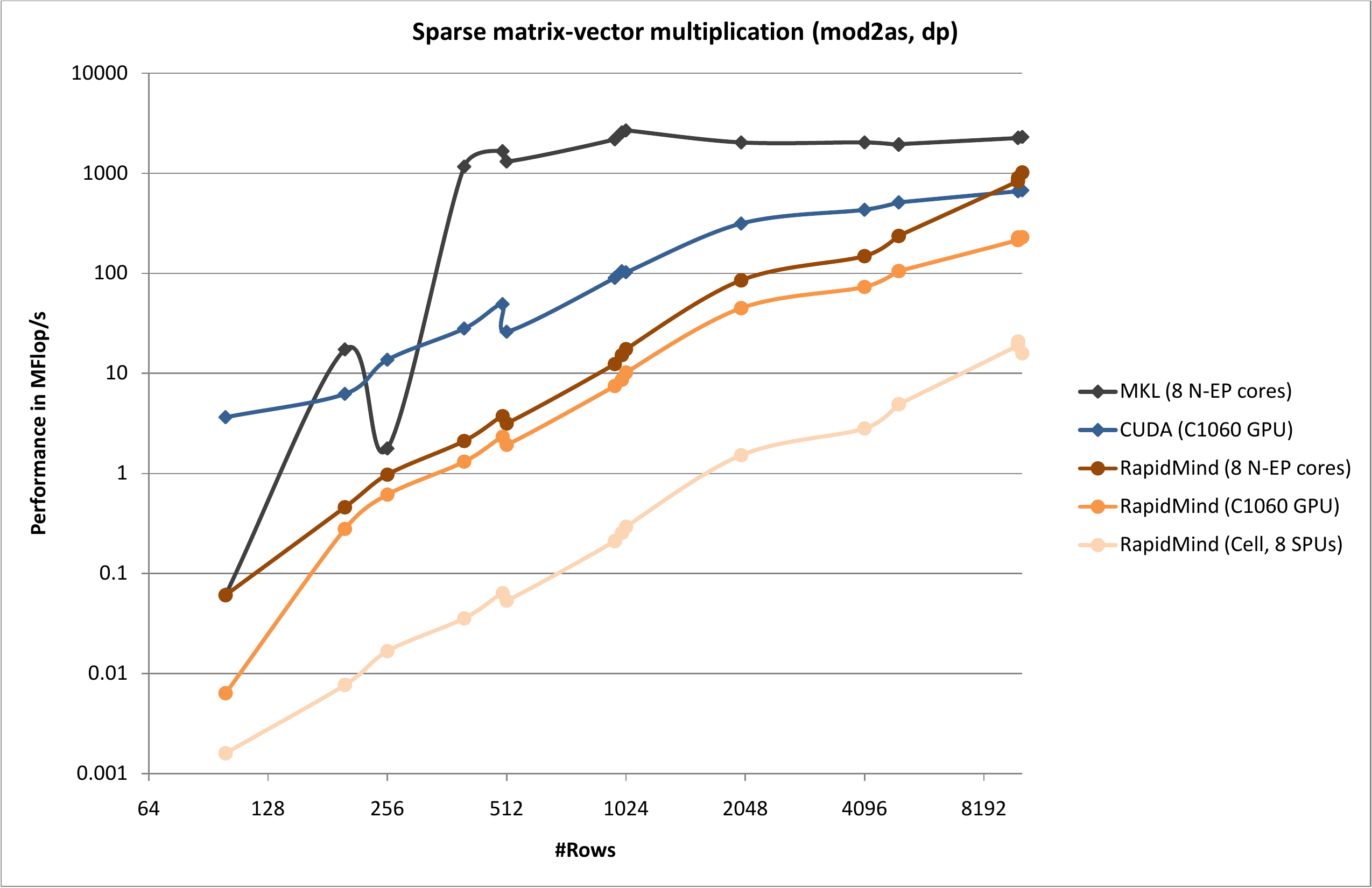}
\end{center}
\caption{Performance comparison of the sparse matrix-vector multiplication. Performance results for the RapidMind implementation on various backends are given and compared with implementations in CUDA and based on MKL. The difference between the MKL and the RapidMind x86-results is less than a factor of 3 for big matrix sizes.}
\label{fig:mod2as}
\end{figure}

Sparse linear algebra is another building block of many scientific algorithms. The sparse matrix-vector multiplication exposes a low computational intensity and is usually memory bound. It is a good example for code that will not perform well on recent hardware accelerators on which the transfer between the x86 host memory and the accelerator memory is a severe bottleneck. Even x86 hardware will only run at a small percentage of its theoretical peak performance. While mod2am reaches more than 90\% of peak, mod2as runs at rates less than 5\% of peak on Nehalem-EP. 
Since this algorithm is not well suited for accelerators, we provided only one RapidMind mod2as implementation and put a special focus on the performance achieved with the x86 backend on Nehalem-EP (shown in Fig.~\ref{fig:mod2as}).
 

The implementation of mod2as is based on~\cite{sparse}. The input matrix $A$ of mod2as is stored in a 3-array variation of the CSR (compressed sparse row) format which can be easily transferred to RapidMind.
The array \texttt{matvals} contains the non-zero elements of $A$, the element \texttt{i} of the integer array \texttt{indx} is the number of the column in $A$ that contains the \texttt{i}-th value in the \texttt{matvals} array and element \texttt{j} of the  integer array \texttt{rowp}
gives the index of the element in the \texttt{matvals} array that is the first non-zero
element in row \texttt{j} of $A$. The input and output vectors are declared as:

\begin{verbatim}
Array<1,Value1i> indx(nelmts);
Array<1,Value1i> rowp(nrows+1);
Array<1,Value1f> matvals(nelmts);

Array<1,Value1f> invec(ncols);
Array<1,Value1f> outvec(nrows);
\end{verbatim}

Once again a helper array based on a call to \texttt{grid(nrows)} is created and used as input vector to allow the correct index computation. The RapidMind program is very clean: 
using RapidMind's \texttt{RM\_FOR()} control structure, the program loops over one row of the input matrix and computes the matrix-vector product.

\begin{verbatim}
Program spMXV = BEGIN {
    In<Value1i> i;  
    Out<Value1f> c;
    
    c = Value1f(0.);
    Value1i j;
    
    RM_FOR(j=rowp[i], j < rowp[i+1] , j++) {
        c += matvals[j] * invec[indx[j]];
    } RM_ENDFOR;
    
} END;
\end{verbatim}

\subsection{One-dimensional Fast Fourier Transformation (mod2f)}

\begin{figure}
\begin{center}
\includegraphics[width=0.80\textwidth]{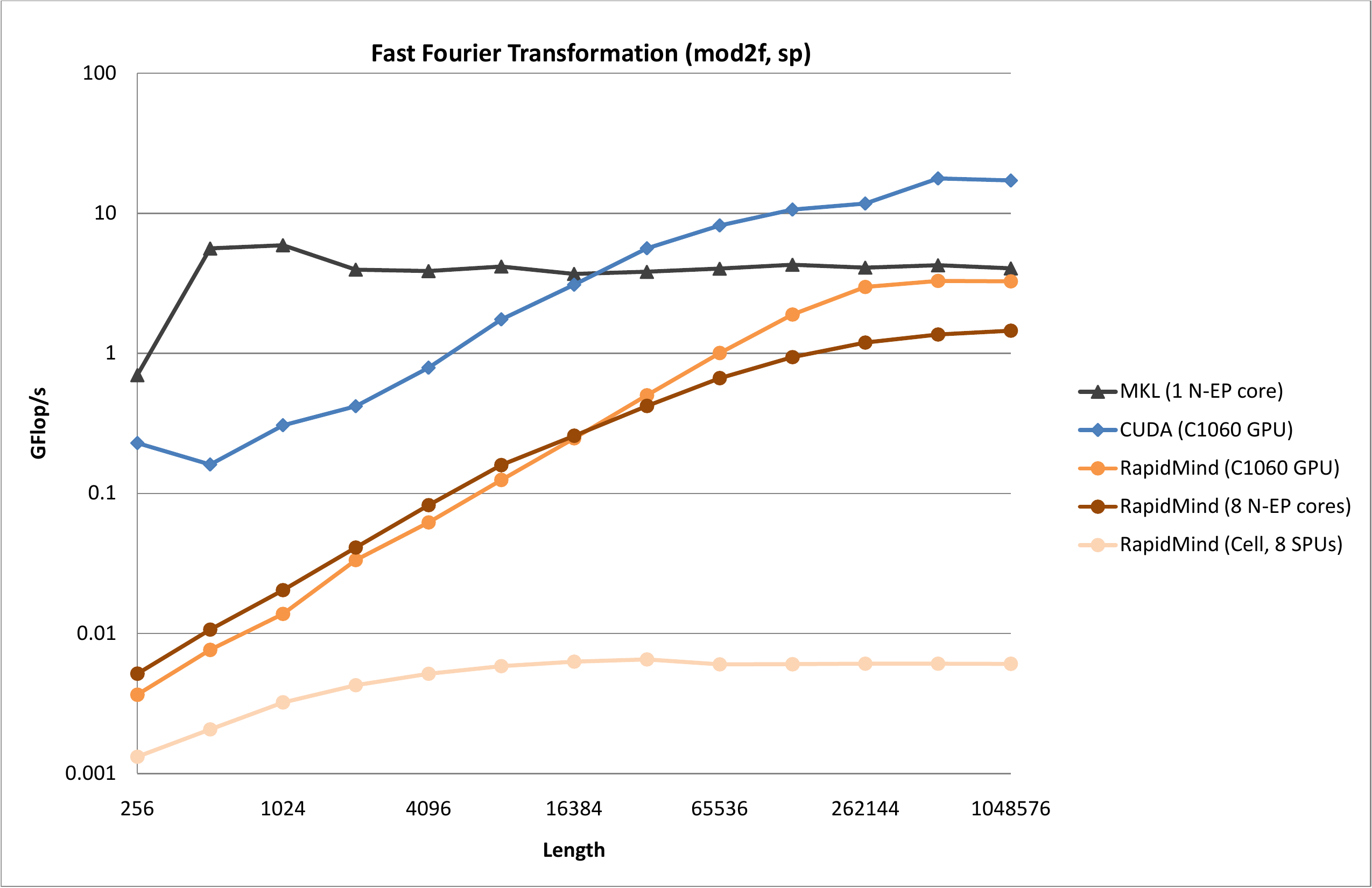}
\end{center}
\caption{Performance comparison of the one-dimensional Fast Fourier Transformation. The RapidMind implementation is compared to a CUDA version of mod2f based on cuFFT and the corresponding MKL implementation.
The gap between the RapidMind cuda-results and the highly-optimized cuFFT version is a factor of 5, the difference between the x86-results and the MKL version is again less than a factor of 3. 
}
\label{fig:mod2f}
\end{figure}

The Fast Fourier Transformation (FFT) is widely used in many scientific programs. Its computational intensity is not as high as for mod2am, but is already in a range where accelerators should be beneficial. The RapidMind version of mod2f computes the FFT using a split-stream algorithm as described in~\cite{FFT}. The implementation is a straightforward conversion of a one butterfly Cooley-Tukey radix-2 FFT; the butterfly kernels are defined as RapidMind programs.
Figure~\ref{fig:mod2f} gives the achieved performance for different platforms and shows that one implementation is able to deliver performance on at least two backends.

\section{Conclusions and Future Work}

The work presented in this paper has shown that RapidMind really offers code portability across various architectures, both multi-core x86 CPUs and accelerators like GPUs or the Cell processor. 
Using RapidMind for the Euroben kernels has been straightforward: the code
development of the first naive implementation took only a few days for each. Adapting the original version to new backends comes practically for free and is a matter of hours and of getting used to the new environments. 

However, performance portability differs: code written naturally without a deep understanding of the hardware and RapidMind's internal mode of operation will not deliver optimal performance in most circumstances and hardly exploit the potential of the hardware. For mod2am, the highly optimized cell-version is able to reach 88\% of the SDK implementation but will deliver poor performance when used on GPUs. The fastest mod2am implementation using CUDA is three times faster than any RapidMind code. For none of the used benchmarks, RapidMind code was able to fully reach the performance of hardware-specific implementations. This is not a big surprise, since it is one of the drawbacks of the achieved code portability. But it is important to state, that the language design optimally supports the compiler. To efficiently use this information to full capacity requires that many people constantly improve all backends, adapting them to the latest hardware and its accompanying language features. 

Recently, RapidMind Inc. has been acquired by Intel. Their product will dissolve in Intel's new language Ct (C for throughput computing)~\cite{Ct}. The basic concepts of both languages have always been very similar.
The acquisition has pros and cons: on one hand, it is up to speculations if -- or when -- Ct will support non-Intel architectures. On the other hand, Intel has much experience with mantaining high-performance compilers and analyzing tools. 

Future work will focus on Intel's Ct and other approaches that are able to deliver support for multiple accelerators. This might include OpenCL, the PGI accelerator compiler, hmpp from CAPS and the StarSs concept. Our work will focus on the question of portability, both in terms of source code and in terms of achievable performance. The number of kernels will be increased to get a better coverage of the ``Berkeley dwarfs''. 


\section*{Acknowledgements}

This work was financially supported by the KONWIHR-II project ``OMI4papps'' and by the PRACE project funded in part by the EU's 7th Framework Programme (FP7/2007-2013) under grant agreement no. RI-211528. We specially thank our colleague Hans Hacker for providing the CUDA ports and performance figures, and JSC and CEA for access to their accelerator systems and support.


\begin{thebibliography}{}  
\bibitem{biba}  The Top500 supercomputing sites, http://www.top500.org/
\bibitem{bib1} 	OpenCL, http://www.khronos.org/opencl/
\bibitem{bib2}	PGI Accelerator Compiler, http://www.pgroup.com/resources/accel.htm
\bibitem{bib3}	CAPS hmpp workbench, www.caps-entreprise.com/hmpp.html
\bibitem{bib4}	J. Planas, R. M. Badia, E. Ayguade, J. Labarta: Hierarchical Task-Based Programming with StarSs, The International Journal of High Performance Computing Applications, Vol. 23, No. 3, 284-299 (2009)
\bibitem{bib12} Sh project, http://libsh.org/
\bibitem{bibFrank} M. Ernst, Ch. Vogelgsang, G. Greiner: Stack Implementation on Programmable Graphics Hardware, VMV 2004: 255-262
\bibitem{bib7}	M. McCool, K. Wadleigh, B. Henderson, H.-Y. Lin: Performance
  evaluation of GPUs using the RapidMind development platform, Proceedings of
  the 2006 ACM/IEEE conference on Supercomputing, 2006

\bibitem{roadrunner}	Los Alamos Lab: Roadrunner, http://www.lanl.gov/roadrunner/
\bibitem{green500}  The Green500 list of energy efficient supercomputers, http://www.green500.org/
\bibitem{euroben}	The Euroben benchmark home page, http://www.euroben.nl/
\bibitem{berkeley} K. Asanovic et. al.: The Landscape of Parallel Computing Research: A View from Berkeley, 2006, http://www.eecs.berkeley.edu/Pubs/TechRpts/2006/EECS-2006-183.pdf
\bibitem{D61} A. Simpson, M. Bull, J. Hill: PRACE Deliverable D6.1 Identification and Categorisation of Applications and Initial Benchmarks Suite, http://www.prace-project.eu/documents/Identification\_and\_Categorisation\_of\_Applications\_and\_Initial\_ Benchmark\_Suite\_final.pdf
\bibitem{D66}	C. Cavazzoni, I. Christadler, G. Erbacci, F. Spiga: PRACE Deliverable D6.6 Report on petascale software libraries and programming models, to appear at http://www.prace-project.eu/documents/public-deliverables-1/
\bibitem{bib11} RapidMind developer site, https://developer.rapidmind.com/sample-code/matrix-multiplication-samples/rm-sgemm-gpu-5938.zip
\bibitem{sparse}	N. Bell, M. Garland: Efficient Sparse Matrix-Vector Multiplication on CUDA, http://www.nvidia.com/object/nvidia\_research\_pub\_001.html
\bibitem{FFT} T. Jansen, B. von Rymon-Lipinski, N. Hanssen, E. Keeve: Fourier Volume Rendering on the GPU Using a Split-Stream-FFT. VMV 2004: 395-403
\bibitem{Ct}	Intel Ct Technology, http://software.intel.com/en-us/data-parallel/








\end{thebibliography}
\end{document}